
%
%
\def\unredoffs{} \def\redoffs{\voffset=-.31truein\hoffset=-.59truein}
\def\speclscape{}
%
%
%
%
\newbox\leftpage \newdimen\fullhsize \newdimen\hstitle \newdimen\hsbody
\tolerance=1000\hfuzz=2pt
\catcode`\@=11 
\def\bigans{b }
\def\answ{b }
%

\ifx\answ\bigans\message{(This will come out unreduced.}
\magnification=1200\unredoffs\baselineskip=.33truein plus 2pt minus 1pt
\hsbody=\hsize \hstitle=\hsize 
\else\message{(This will be reduced.} \let\l@r=L
\magnification=1000\baselineskip=16pt plus 2pt minus 1pt \vsize=7truein
\redoffs \hstitle=8truein\hsbody=4.75truein\fullhsize=10truein\hsize=\hsbody
\output={\ifnum\pageno=0 
  \shipout\vbox{\speclscape{\hsize\fullhsize\makeheadline}
   \hbox to \fullhsize{\hfill\pagebody\hfill}}\advancepageno
  \else
 \almostshipout{\leftline{\vbox{\pagebody\makefootline}}}\advancepageno
  \fi}
\def\almostshipout#1{\if L\l@r \count1=1 \message{[\the\count0.\the\count1]}
      \global\setbox\leftpage=#1 \global\let\l@r=R
 \else \count1=2
  \shipout\vbox{\speclscape{\hsize\fullhsize\makeheadline}
      \hbox to\fullhsize{\box\leftpage\hfil#1}}  \global\let\l@r=L\fi}
\fi
%
\newcount\yearltd\yearltd=\year\advance\yearltd by -1900

\def\Title#1#2{\nopagenumbers\abstractfont\hsize=\hstitle\rightline{#1}%
\vskip 1in\centerline{\titlefont #2}\abstractfont\vskip .5in\pageno=0}
\def\Date#1{\vfill\leftline{#1}\tenpoint\supereject\global\hsize=\hsbody%
\footline={\hss\tenrm\folio\hss}}
%

\def\draftmode{\message{ DRAFTMODE }\def\draftdate{{\rm preliminary draft:
\number\month/\number\day/\number\yearltd\ \ \hourmin}}%
\headline={\hfil\draftdate}\writelabels\baselineskip=20pt plus 2pt minus 2pt
 {\count255=\time\divide\count255 by 60 \xdef\hourmin{\number\count255}
  \multiply\count255 by-60\advance\count255 by\time
  \xdef\hourmin{\hourmin:\ifnum\count255<10 0\fi\the\count255}}}
\def\nolabels{\def\wrlabeL##1{}\def\eqlabeL##1{}\def\reflabeL##1{}}
\def\writelabels{\def\wrlabeL##1{\leavevmode\vadjust{\rlap{\smash%
{\line{{\escapechar=` \hfill\rlap{\sevenrm\hskip.03in\string##1}}}}}}}%
\def\eqlabeL##1{{\escapechar-1\rlap{\sevenrm\hskip.05in\string##1}}}%
\def\reflabeL##1{\noexpand\llap{\noexpand\sevenrm\string\string\string##1}}}
\nolabels
%
\global\newcount\secno \global\secno=0
\global\newcount\meqno \global\meqno=1
\def\newsec#1{\global\advance\secno by1\message{(\the\secno. #1)}
\global\subsecno=0\eqnres@t\noindent{\bf\the\secno. #1}
\writetoca{{\secsym} {#1}}\par\nobreak\medskip\nobreak}
\def\eqnres@t{\xdef\secsym{\the\secno.}\global\meqno=1\bigbreak\bigskip}
\def\sequentialequations{\def\eqnres@t{\bigbreak}}\xdef\secsym{}
\global\newcount\subsecno \global\subsecno=0
\def\subsec#1{\global\advance\subsecno by1\message{(\secsym\the\subsecno. #1)}
\ifnum\lastpenalty>9000\else\bigbreak\fi
\noindent{\it\secsym\the\subsecno. #1}\writetoca{\string\quad
{\secsym\the\subsecno.} {#1}}\par\nobreak\medskip\nobreak}
\def\appendix#1#2{\global\meqno=1\global\subsecno=0\xdef\secsym{\hbox{#1.}}
\bigbreak\bigskip\noindent{\bf Appendix #1. #2}\message{(#1. #2)}
\writetoca{Appendix {#1.} {#2}}\par\nobreak\medskip\nobreak}
%
%
\def\eqnn#1{\xdef #1{(\secsym\the\meqno)}\writedef{#1\leftbracket#1}%
\global\advance\meqno by1\wrlabeL#1}
\def\eqna#1{\xdef #1##1{\hbox{$(\secsym\the\meqno##1)$}}
\writedef{#1\numbersign1\leftbracket#1{\numbersign1}}%
\global\advance\meqno by1\wrlabeL{#1$\{\}$}}
\def\eqn#1#2{\xdef #1{(\secsym\the\meqno)}\writedef{#1\leftbracket#1}%
\global\advance\meqno by1$$#2\eqno#1\eqlabeL#1$$}
%
\newskip\footskip\footskip14pt plus 1pt minus 1pt 
\def\footnotefont{\ninepoint}\def\f@t#1{\footnotefont #1\@foot}
\def\f@@t{\baselineskip\footskip\bgroup\footnotefont\aftergroup\@foot\let\next}
\setbox\strutbox=\hbox{\vrule height9.5pt depth4.5pt width0pt}
\global\newcount\ftno \global\ftno=0
\def\foot{\global\advance\ftno by1\footnote{$^{\the\ftno}$}}
%
\newwrite\ftfile
\def\footend{\def\foot{\global\advance\ftno by1\chardef\wfile=\ftfile
$^{\the\ftno}$\ifnum\ftno=1\immediate\openout\ftfile=foots.tmp\fi%
\immediate\write\ftfile{\noexpand\smallskip%
\noexpand\item{f\the\ftno:\ }\pctsign}\findarg}%
\def\footatend{\vfill\eject\immediate\closeout\ftfile{\parindent=20pt
\centerline{\bf Footnotes}\nobreak\bigskip\input foots.tmp }}}
\def\footatend{}
%
%
\global\newcount\refno \global\refno=1
\newwrite\rfile
\def\ref{[\the\refno]\nref}
\def\nref#1{\xdef#1{[\the\refno]}\writedef{#1\leftbracket#1}%
\ifnum\refno=1\immediate\openout\rfile=refs.tmp\fi
\global\advance\refno by1\chardef\wfile=\rfile\immediate
\write\rfile{\noexpand\item{#1\ }\reflabeL{#1\hskip.31in}\pctsign}\findarg}
\def\findarg#1#{\begingroup\obeylines\newlinechar=`\^^M\pass@rg}
{\obeylines\gdef\pass@rg#1{\writ@line\relax #1^^M\hbox{}^^M}%
\gdef\writ@line#1^^M{\expandafter\toks0\expandafter{\striprel@x #1}%
\edef\next{\the\toks0}\ifx\next\em@rk\let\next=\endgroup\else\ifx\next\empty%
\else\immediate\write\wfile{\the\toks0}\fi\let\next=\writ@line\fi\next\relax}}
\def\striprel@x#1{} \def\em@rk{\hbox{}}
\def\lref{\begingroup\obeylines\lr@f}
\def\lr@f#1#2{\gdef#1{\ref#1{#2}}\endgroup\unskip}

\def\addref#1{\immediate\write\rfile{\noexpand\item{}#1}} 
\def\startrefs#1{\immediate\openout\rfile=refs.tmp\refno=#1}
\def\xref{\expandafter\xr@f}\def\xr@f[#1]{#1}
\def\refs#1{\count255=1[\r@fs #1{\hbox{}}]}
\def\r@fs#1{\ifx\und@fined#1\message{reflabel \string#1 is undefined.}%
\nref#1{need to supply reference \string#1.}\fi%
\vphantom{\hphantom{#1}}\edef\next{#1}\ifx\next\em@rk\def\next{}%
\else\ifx\next#1\ifodd\count255\relax\xref#1\count255=0\fi%
\else#1\count255=1\fi\let\next=\r@fs\fi\next}
%

%
\newwrite\ffile\global\newcount\figno \global\figno=1
\def\fig{Figure~\the\figno\nfig}
\def\nfig#1{\xdef#1{Figure~\the\figno}%
\writedef{#1\leftbracket fig.\noexpand~\the\figno}%
\ifnum\figno=1\immediate\openout\ffile=figs.tmp\fi\chardef\wfile=\ffile%
\immediate\write\ffile{\noexpand\medskip\noexpand\item{Fig.\ \the\figno. }
\reflabeL{#1\hskip.55in}\pctsign}\global\advance\figno by1\findarg}
\def\xfig{\expandafter\xf@g}\def\xf@g fig.\penalty\@M\ {}
\def\figs#1{figs.~\f@gs #1{\hbox{}}}
\def\f@gs#1{\edef\next{#1}\ifx\next\em@rk\def\next{}\else
\ifx\next#1\xfig #1\else#1\fi\let\next=\f@gs\fi\next}
\newwrite\lfile
{\escapechar-1\xdef\pctsign{\string\%}\xdef\leftbracket{\string\{}
\xdef\rightbracket{\string\}}\xdef\numbersign{\string\#}}

\def\writestop{\def\writestoppt{\immediate\write\lfile{\string\pageno%
\the\pageno\string\startrefs\leftbracket\the\refno\rightbracket%
\string\def\string\secsym\leftbracket\secsym\rightbracket%
\string\secno\the\secno\string\meqno\the\meqno}\immediate\closeout\lfile}}
\def\writestoppt{}\def\writedef#1{}
\def\seclab#1{\xdef #1{\the\secno}\writedef{#1\leftbracket#1}\wrlabeL{#1=#1}}
\def\subseclab#1{\xdef #1{\secsym\the\subsecno}%
\writedef{#1\leftbracket#1}\wrlabeL{#1=#1}}
\newwrite\tfile \def\writetoca#1{}
\def\leaderfill{\leaders\hbox to 1em{\hss.\hss}\hfill}
\def\writetoc{\immediate\openout\tfile=toc.tmp
   \def\writetoca##1{{\edef\next{\write\tfile{\noindent ##1
   \string\leaderfill {\noexpand\number\pageno} \par}}\next}}}

%
\catcode`\@=12 
%
\edef\tfontsize{\ifx\answ\bigans scaled\magstep3\else scaled\magstep4\fi}
\font\titlerm=cmr10 \tfontsize \font\titlerms=cmr7 \tfontsize
\font\titlermss=cmr5 \tfontsize \font\titlei=cmmi10 \tfontsize
\font\titleis=cmmi7 \tfontsize \font\titleiss=cmmi5 \tfontsize
\font\titlesy=cmsy10 \tfontsize \font\titlesys=cmsy7 \tfontsize
\font\titlesyss=cmsy5 \tfontsize \font\titleit=cmti10 \tfontsize
\skewchar\titlei='177 \skewchar\titleis='177 \skewchar\titleiss='177
\skewchar\titlesy='60 \skewchar\titlesys='60 \skewchar\titlesyss='60
\def\titlefont{\def\rm{\fam0\titlerm}
\textfont0=\titlerm \scriptfont0=\titlerms \scriptscriptfont0=\titlermss
\textfont1=\titlei \scriptfont1=\titleis \scriptscriptfont1=\titleiss
\textfont2=\titlesy \scriptfont2=\titlesys \scriptscriptfont2=\titlesyss
\textfont\itfam=\titleit \def\it{\fam\itfam\titleit}\rm}
 \ifx\answ\bigans\else scaled\magstep1\fi
\ifx\answ\bigans\def\abstractfont{\tenpoint}\else
\font\abssl=cmsl10 scaled \magstep1
\font\absrm=cmr10 scaled\magstep1 \font\absrms=cmr7 scaled\magstep1
\font\absrmss=cmr5 scaled\magstep1 \font\absi=cmmi10 scaled\magstep1
\font\absis=cmmi7 scaled\magstep1 \font\absiss=cmmi5 scaled\magstep1
\font\abssy=cmsy10 scaled\magstep1 \font\abssys=cmsy7 scaled\magstep1
\font\abssyss=cmsy5 scaled\magstep1 \font\absbf=cmbx10 scaled\magstep1
\skewchar\absi='177 \skewchar\absis='177 \skewchar\absiss='177
\skewchar\abssy='60 \skewchar\abssys='60 \skewchar\abssyss='60
\def\abstractfont{\def\rm{\fam0\absrm}
\textfont0=\absrm \scriptfont0=\absrms \scriptscriptfont0=\absrmss
\textfont1=\absi \scriptfont1=\absis \scriptscriptfont1=\absiss
\textfont2=\abssy \scriptfont2=\abssys \scriptscriptfont2=\abssyss
\textfont\itfam=\bigit \def\it{\fam\itfam\bigit}\def\footnotefont{\tenpoint}%
\textfont\slfam=\abssl \def\sl{\fam\slfam\abssl}%
\textfont\bffam=\absbf \def\bf{\fam\bffam\absbf}\rm}\fi
\def\tenpoint{\def\rm{\fam0\tenrm}
\textfont0=\tenrm \scriptfont0=\sevenrm \scriptscriptfont0=\fiverm
\textfont1=\teni  \scriptfont1=\seveni  \scriptscriptfont1=\fivei
\textfont2=\tensy \scriptfont2=\sevensy \scriptscriptfont2=\fivesy
\textfont\itfam=\tenit \def\it{\fam\itfam\tenit}\def\footnotefont{\ninepoint}%
\textfont\bffam=\tenbf \def\bf{\fam\bffam\tenbf}\def\sl{\fam\slfam\tensl}\rm}
\font\ninerm=cmr9 \font\sixrm=cmr6 \font\ninei=cmmi9 \font\sixi=cmmi6
\font\ninesy=cmsy9 \font\sixsy=cmsy6 \font\ninebf=cmbx9
\font\nineit=cmti9 \font\ninesl=cmsl9 \skewchar\ninei='177
\skewchar\sixi='177 \skewchar\ninesy='60 \skewchar\sixsy='60
\def\ninepoint{\def\rm{\fam0\ninerm}
\textfont0=\ninerm \scriptfont0=\sixrm \scriptscriptfont0=\fiverm
\textfont1=\ninei \scriptfont1=\sixi \scriptscriptfont1=\fivei
\textfont2=\ninesy \scriptfont2=\sixsy \scriptscriptfont2=\fivesy
\textfont\itfam=\ninei \def\it{\fam\itfam\nineit}\def\sl{\fam\slfam\ninesl}%
\textfont\bffam=\ninebf \def\bf{\fam\bffam\ninebf}\rm}
%
%

\hyphenation{anom-aly anom-alies coun-ter-term coun-ter-terms}
\def\inv{^{\raise.15ex\hbox{${\scriptscriptstyle -}$}\kern-.05em 1}}

\def\Dsl{\,\raise.15ex\hbox{/}\mkern-13.5mu D} 
\def\dsl{\raise.15ex\hbox{/}\kern-.57em\partial}

\font\bigit=cmti10 scaled \magstep1
\def\lspace{\ifx\answ\bigans{}\else\qquad\fi}
\def\lbspace{\ifx\answ\bigans{}\else\hskip-.2in\fi} 
\def\boxeqn#1{\vcenter{\vbox{\hrule\hbox{\vrule\kern3pt\vbox{\kern3pt
	\hbox{${\displaystyle #1}$}\kern3pt}\kern3pt\vrule}\hrule}}}
\def\mbox#1#2{\vcenter{\hrule \hbox{\vrule height#2in
		\kern#1in \vrule} \hrule}}  
%

\def\grad#1{\,\nabla\!_{{#1}}\,}

\def\darr#1{\raise1.5ex\hbox{$\leftrightarrow$}\mkern-16.5mu #1}

\def\roughly#1{\raise.3ex\hbox{$#1$\kern-.75em\lower1ex\hbox{$\sim$}}}

\def\tiny{\scriptscriptstyle\rm\!\!}
\def\div{\nabla_{\tiny\!\perp}\!\cdot\!}
\def\dz{\partial_z}
\def\curl{\nabla_{\tiny\!\perp}\!\times\!}
\def\kbT{k_{\scriptscriptstyle\rm B}T}

\def\bo#1{{\cal O}(#1)}
\def\angstrom{{\hbox{\AA}}}
\def\grad{\nabla_{\tiny\! \perp}}
\def\cross{\!\!\times\!\!}
\def\dnb{\delta {\vec n}}
\def\dot{\!\cdot\!}
\def\identical{\equiv}
\lref\GP{G.~Grinstein and R.~Pelcovits, Phys. Rev. Lett.
{\bf 47}, 856 (1981); Phys. Rev. A {\bf 26}, 915 (1982).}
\lref\TONER{J.~Toner, Phys. Rev. Lett. {\bf 68}, 1331 (1992).}
\lref\KLN{P.~Le Doussal and D.R.~Nelson, Europhys. Lett. {\bf 15}, 161 (1991);
R.D. Kamien, P. Le~Doussal, and D.R.~Nelson, Phys. Rev. A {\bf
45}, 8727 (1992); Phys. Rev. E {\bf 48},
4116 (1993).}
\lref\KL{R.D.~Kamien and T.C.~Lubensky, J. Phys. I (Paris) {\bf 3} 2131
(1993).}
\lref\PN{P.~Nelson and T.~Powers, Phys. Rev. Lett. {\bf 69}, 3409 (1992);
J. Phys. II (Paris) {\bf 3} 1535 (1993).}
\lref\SB{J.V.~Selinger and R.F.~Bruinsma, Phys. Rev. A {\bf 43}, 2910
(1991).}
\lref\PRP{D.C.~Rau, B.~Lee and V.A.~Parsegian,
Proc. Natl. Acad. Sci. USA {\bf 81}, 2621 (1984);
R.~Podgornik, D.C.~Rau and V.A.~Parsegian,
Macromolecules {\bf 22}, 1780 (1989).}
\lref\TARME{V.G.~Taratura and R.B.~Meyer, Liquid Crystals {\bf 2} 373
(1987).}
\lref\MEYER{R.B.~Meyer, {\sl Polymer Liquid
Crystals}, edited by A. Ciferri, W.R. Kringbaum and R.B. Meyer (Academic, New
York, 1982) Chapter 6.}
\lref\KNT{R.D.~Kamien, D.R.~Nelson and J.~Toner, unpublished (1994).}
\lref\MEYERa{R.B.~Meyer, Appl. Phys. Lett. {\bf 12}, 281 (1968);
Appl. Phys. Lett. {\bf 14}, 208 (1969).}
\lref\KN{R.D.~Kamien and D.R.~Nelson, Institute for Advanced Study
Preprint IASSNS-HEP-94/33, (1994).}
\lref\POD{V.A.~Parsegian, R.~Podgornik, D.C.~Rau and H.~Strey,
private communication.}
\Title{IASSNS-HEP-94/48}{
Anomalous Elasticity of Polymer Cholesterics}
\centerline{Randall D. Kamien\foot{email: kamien@guinness.ias.edu}}
\smallskip
\centerline{\sl School of Natural Sciences, Institute for Advanced Study,
Princeton, NJ 08540}
\smallskip\centerline{and}\smallskip\centerline{John Toner\foot{email:
toner@watson.ibm.com}}
\smallskip
\centerline{\sl IBM Research Division, Yorktown Heights, NY 10598}
\vskip .3in
We show that polymer cholesterics have much longer pitches than
comparable short molecule cholesterics, due to their
anomalous elasticity. The pitch $P$ of
a chiral mixture with concentration $c$ near the racemic (non-chiral)
concentration
$c^*$ diverges like
$\vert c-c^*\vert^{-\nu}$
with,
according to our very precise calculations \ref\POIN{That is, a
two-loop $4-\epsilon$ expansion; see also
R.~Poindexter and
F.T.~Cat, unpublished.},
$\nu=1.43 \pm 0.04$ (for short molecule cholesterics $\nu=1$).
The short molecule law is recovered for polymers of finite molecular
length $\ell$ once the pitch is longer than a length that diverges like
$\ell^\gamma$ with $\gamma=0.67 \pm 0.01$. Our predictions could
be tested by measurements of the pitch in DNA \POD .
\Date{12 August 1994; revised 17 August 1994}
Nature is chiral. In
short-chain chiral molecules this leads to
structures
so numerous that we must borrow words from Greek in order to
categorize them properly.
The simplest of these structures is the cholesteric, in which the
molecules lie perpendicular to a ``pitch'' axis, about which they rotate
as one moves through space along it.

The standard mean field theory of cholesterics
\ref\PLC{P.G.~de Gennes and J.~Prost, {\sl The Physics of Liquid Crystals},
Second Edition, Chap. VII, (Oxford University Press, New York, 1993).}\
predicts that the
cholesteric pitch $P$ diverges according to
$P\propto \vert c-c^*\vert^{-1},$
where $c$ is the concentration of some chiral additive and $c^*$ the
concentration of the racemic ({\sl i.e.}, non-chiral) mixture.

In short molecule cholesterics, this mean field theory is exactly
correct; all effects of fluctuations can be absorbed into perfectly
finite renormalizations of the effective parameters of the mean-field
theory. However, this is certain {\sl not} to be the case for polymeric
cholesterics, since it has been shown \TONER\ that, as in smectics \GP ,
the effects of
thermal fluctuations on polymer {\sl nematics}
lead to infinite renormalizations
of the elastic constants (for infinitely long
polymers).
This is known as ``anomalous
elasticity''.

Here we investigate the effect of anomalous
elasticity on polymer cholesterics, of which DNA is a classic example
\ref\WATCRIK{J.D.~Watson and F.H.C.~Crick, Nature {\bf 171}:737 (1953).}.
We find that
once the pitch predicted by mean field theory ($P_0$) is much greater
than the intrinsic length $\xi^N_\perp$ beyond which the elasticity
becomes anomalous, the {\sl actual} pitch $P$ is given by:
\eqn\pitpit{P = \xi^N_\perp \left( {P_0 \over
\xi^N_\perp}\right)^\nu\times\bo{1},}
with
$\nu=1.43 \pm 0.04$.
This rapidly diverging pitch reflects a sort of
fluctuation suppression of the bare chirality, a phenomenon that
occurs \PN\ in chiral {\sl membranes} as well, although the
effect found here is quantitatively much stronger than in chiral
membranes, being an algebraically diverging correction in our case, in
contrast to a logarithmic correction in membranes.
For $P_0<<\xi^N_\perp$, $P=P_0$.
For polymers whose
interactions are primarily steric entropic ({\sl i.e.}, due to their
meandering and bumping into each other),
reference \TONER\ showed that $\xi^N_\perp = {L^{4/3}_P /
a^{1/3}}$, where $L_P \equiv {\kappa /\kbT }$ is the
orientational persistence length for a single isolated polymer, (with
$\kappa$ the polymer bend modulus), and $a$ is the mean polymer
spacing.

In addition, we consider the effects of a finite (but long) polymer
length on {\sl both} polymer nematics {\sl and} cholesterics. We find
that these were incorrectly treated in \TONER .
Here, we determine the
correct behavior of the renormalized Frank elastic constants as a
function of polymer length $\ell$
for polymer nematics, which proves to be:
\eqn\Flen{K_1 = \rho_0\kbT\ell,
\qquad K_{2,3} = K^0_{2,3} \left({\ell \over \ell_0}\right)^{\phi_{2,3}},}
where $\phi_2 = 0.20\pm 0.01$
and $\phi_3 = 0.15 \pm 0.02$, respectively.
Here,
\eqn\theellone{\ell_0 \identical {\left[K^0_2 (K^0_3)^3\right]^{1/2} a^2
\xi^N_\perp\over
(\kbT)^2},}
with $K^0_{2,3}$ the bare Frank constants and $a$ the mean
polymer spacing. In addition, $\rho_0$ is the mean polymer density.
For entropic steric polymers as described earlier, $\ell_0 = {L_P^3/
a^2}$.
The value of $K_1$ was originally
predicted by Meyer \MEYER .

For polymer cholesterics, this finite polymer length cutoff
leads to a crossover back to the mean field
law equation $P\propto P_0$ once the {\sl actual} pitch $P$
exceeds a length
\eqn\xil{\xi_\perp^L(\ell) =\xi^N_\perp \left({\ell \over
\ell_0}\right)^\gamma}
where $\gamma=0.67 \pm 0.01.$
Mean field theory \PLC\ predicts that the pitch is
proportional to $K_2$ for fixed chirality; hence, in this large pitch
($P>>\xi^L_\perp (\ell)$)
regime, the pitch becomes $\ell$ dependent, diverging in the same way
with $\ell$ as $K_2$; {\sl i.e.},
\eqn\lpitch{P(\ell) = P_0\left({\ell \over \ell_0}\right)^{\phi_2}.}
This surprising prediction could be tested
experimentally in DNA by using enzymes to cut
the DNA while keeping other microscopic properties fixed.

The above predictions for the behavior of the pitch in a polymer
cholesteric with long, but finite polymers, can be summarized
for $P_0>>\xi^N_\perp$ by the scaling law
\eqn\escalinglaw{P=P_0^\nu h\left[{P_0\over P^c_0(\ell)}\right],}
where $h(x\rightarrow 0)\rightarrow
\left(\xi^N_\perp\right)^{1-\nu}\times\bo{1}$,
$P^c_0(\ell) =\xi^N_\perp\left(\xi^L_\perp(\ell)/
\xi^N_\perp\right)^{(1/\nu)}$,
and $h(x\rightarrow\infty)\rightarrow \left(\xi^N_\perp x\right)^{1-\nu}$.
Our full predictions for the pitch $P$ are plotted in \fig\fone
{Pitch $P$ as a function of the bare pitch $P_0$.  For
actual pitches $P$ below $\xi^N_\perp$, harmonic elasticity theory applies.
For larger pitches, anharmonic effects become important, leading to
a nonlinear dependence of $P$ on $P_0$.  For pitches longer
that $\xi^L_\perp$, the finite length of the polymers cuts off the
anharmonic effects, and we return to a linear relation again, with
a new slope.}.

The derivation of the above results begins
by formulating the fully rotationally invariant, anharmonic
theory
of a polymer in a liquid crystal matrix.  After \KLN , we expect that at
long distances, this will
also describe polymer liquid crystals in an isotropic solvent.

The free energy has four terms:
1) A term aligning the polymers with the surrounding nematic matrix.
2) The Frank free energy for distortions of this surrounding matrix.
3) The entropy of mixing for polymer ends.
4) A chiral term which generates spontaneous twist.
In addition, as emphasized in past
treatments \refs{\TARME,\KLN,\TONER}, polymer nematics
are distinguished from short-molecule nematics by a constraint relating
the polymer density to the local polymer tangent field ${\vec m}$,
defined to point parallel to the local unit polymer tangent
with magnitude equal to the polymer areal density in the plane
perpendicular to ${\vec m}$.

In terms of $\vec m$ and the usual
nematic director $\hat n$, the polymer-nematic coupling is \KL
\eqn\efree{F_{\rm pol}={1\over 2}\int d^3\!x\, I_{\mu\nu}(m_\mu -\rho_0n_\mu)
(m_\nu- \rho_0n_\nu)}
where $I_{\mu\nu}$ is
diagonal and $I_{xx}=I_{yy}$
(here and hereafter, $\hat z$ is the average polymer direction).
Defining the vector $\vec t$ in the $xy$-plane via
$\vec m = \rho\hat z + \vec t$, $t_z=0$ and
writing $\vec n =  \left[\delta n_x, \delta n_y,
\sqrt{1-(\delta\vec n)^2}\right]$,
\efree\ becomes
\eqn\efreeii{F_{\rm pol}={1\over 2}
\int d^3\!x\,\left\{{E\over\rho_0^2}
\left(\rho - \rho_0\sqrt{1-(\delta\vec n)^2}\right)^2
+B(\vec t - \rho_0\delta \vec n)^2\right\},}
where $E \identical I_{zz}\rho_0^2$ and $B \identical I_{xx}$.
It was shown in \TONER\ that in the infinite polymer case the only
relevant anharmonic
term comes from the second order term in the expansion of
$\sqrt{1-\dnb^2}$.

The only relevant
terms \TONER\ in the Frank free energy are
\eqn\efrank{F_{\vec n} = {1\over 2}\int d^3\!x\,
\left\{K_1(\grad\dot\dnb)^2 + K_2(\grad\cross\dnb)^2 +K_3(\partial_z\dnb)^2
\right\}.}

The entropy of mixing of polymer heads and tails, in the limit of long
polymers (where the density of heads and tails is low) should be just
that of an ideal solution; hence the contribution to the free energy is
just $\kbT$ times this entropy \KLN :
$f=\kbT\rho_H\ln\rho_H
+\kbT\rho_T\ln\rho_T$.
Expanding this about the equilibrium density
$\rho_0=\rho_H\ell$ gives
\eqn\eends{F_{\rm ends}= {1\over 2}\int d^3\!x\,G(\rho_H-\rho_T)^2}
where $G=\partial^2f/\partial(\rho_H-\rho_T)^2=
\kbT
(\langle\rho_H\rangle^{-1}+\langle\rho_T\rangle^{-1})/4
=\kbT\ell/2\rho_0$.

Now we turn to the constraint.
Conservation of polymer requires
$\nabla\!\cdot\!\vec m = \rho_H - \rho_T$.
This becomes, to leading order
\eqn\econs{\partial_z\rho + \grad\dot\vec t = \rho_H-\rho_T . }
Note that as $\ell\rightarrow\infty$, the
quadratic term in \eends\ forces the constraint $\nabla\dot\vec m=0$ to
hold everywhere, as expected.

Following \refs{\TARME,\KLN,\TONER}, we introduce
a phonon-like
field through
${\vec t =\rho_0\partial_z\vec u}$, and
add
an ``incompatibility field'' $\omega$ defined by
${\delta\rho=-\rho_0\grad\dot\vec u+ \omega}$. Using these definitions
and equation \econs\ to write $\rho_H - \rho_T$ in terms of $\vec u$ and
$\omega$, and integrating out $\dnb$, we
are led to the full free energy:
\eqn\efreefull{F={1\over 2}\int d^3\!x\,\left\{E\left[ \div\vec u +
{1\over 2}(\partial_z\vec u)^2 +\omega\right]^2 + G(
\partial_z\omega)^2\right\} + F_{\vec n}[\dz\vec u] .}

In the limit $\ell \rightarrow \infty$, $G \rightarrow \infty$,
and $\omega$
is forced to be independent of $z$.
In this situation, $\omega$ can be eliminated by a change of variables
$\vec u \equiv \vec u' + \grad\Omega(x,y)$, where $\grad^2\Omega=-\omega$,
and equation \efreefull\ reduces to precisely
the model
considered by \TONER . It was shown there that, for nematic polymers
below four dimensions (like most experimental systems), thermal
fluctuations {\sl always} invalidate harmonic elastic theory at
sufficiently long length scales. Harmonic elastic theory only applies
when the length scale under consideration in the $xy$-plane
is less than
$\xi^N_\perp$, or that in the z-direction is less than $\xi^N_z$, where
\eqn\enonamea{\xi^N_\perp =
\left({(K_2^0)^{3/2}(K_3^0)^{1/2}\over
E_0\kbT}\right)^{1/(4-d)},\qquad\quad  \xi^N_z = \xi^N_\perp
\sqrt{K_3^0\over K_2^0}}
where $K_2^0$, $K_3^0$ and $E_0$ are the bare values of
the elastic
constants, and we've generalized from 3 to d dimensions.
For polymers that interact primarily through steric repulsion
({\sl i.e.}, by bumping into each other as they meander),
this becomes \SB ,
in $d=3$ (the physically relevant dimension
\ref\ewitten{For an alternative view,
see M.B.~Green, J.H.~Schwarz and E.~Witten, {\sl Superstring Theory},
(Cambridge University Press, Cambridge, 1987).}),
$\xi^N_\perp = P^{4/3}/a^{1/3}$ and
$\xi^N_z = P^{5/3}/a^{2/3}$.

For length scales longer than  $\xi^N_\perp$ and $\xi^N_z$ in the appropriate
directions,
the equilibrium linear response and correlation functions of a nematic
polymer are characterized by {\sl wavevector dependent} elastic moduli
$E(\vec q)$, $K_2(\vec q)$, and $K_3(\vec q)$, given by
\eqn\eanom{\eqalign{K_{2,3}(\vec q) &=
(q_z\xi^N_z)^{-\eta_{2,3}}f_{2,3}\left[{\left(q_z\xi^N_z\right)^\zeta
\over q_\perp\xi^N_\perp}\right]
\propto\left\{\eqalign{&q_z^{-\eta_{2,3}},\qquad\;\,\,\,
(q_z\xi^N_z)^\zeta >> q_\perp\xi^N_\perp\cr
&q_\perp^{-\eta_{2,3}/\zeta},\qquad
(q_z\xi^N_z)^\zeta << q_\perp\xi^N_\perp\cr
}\right.\cr&\cr
E(\vec q) &= (q_z\xi^N_z)^{\eta_\perp}
f_E\left[{\left(q_z\xi^N_z\right)^\zeta
\over q_\perp\xi^N_\perp}\right]
\propto\left\{\eqalign{&q_z^{\eta_\perp},
\qquad\;\,\,\,(q_z\xi^N_z)^\zeta>> q_\perp\xi^N_\perp\cr
&q_\perp^{\eta_\perp/\zeta},\qquad (q_z\xi^N_z)^\zeta <<
q_\perp\xi^N_\perp\cr}\right.\cr}}
where $f_{E,2,3}$ are universal scaling functions, and the universal
exponents $\zeta$, $\eta_\perp$, $\eta_2$, and $\eta_3$ satisfy two
exact scaling relations, which, in $d=3$, read $\eta_\perp + \eta_2 +
\eta_3 = 1$ and $\zeta=1+{\eta_2-\eta_3 \over 2}$. This behavior is
what we mean by ``anomalous elasticity''; the usual, constant elastic
modulus behavior we will refer to as ``conventional elasticity'',
with no pejorative condescension intended.
In reference \TONER ,
the numerical values of $\eta_2$ and $\eta_3$ were calculated from an
$\epsilon=4-d$ expansion; the best resulting numerical values in $d=3$
\ref\mess{This corrects small numerical errrors in \TONER .}
are $\eta_2=0.31 \pm 0.02$ and $\eta_3=0.24 \pm 0.02$. Henceforth
numerical values and error bars for all
other exponents will be obtained by deriving {\sl exact} expressions
for them in terms of $\eta_2$ and $\eta_3$, and then
using the just-quoted numerical values and error bars in those exact
expressions.

When the length $\ell$ of the polymers, and hence the elastic constant
$G$ in \efreefull\ are finite, this behavior ultimately gets cut off at
sufficiently long length scales, beyond which the conventional
({\sl i.e.}, non-anomalous)
elastic behavior of a conventional nematic is again recovered. To
calculate this length, it is first instructive to consider the harmonic
approximation, in which we can integrate out
$\omega$.  We thereby obtain, in Fourier space,
\eqn\efinitelength{F={1\over 2}\int d^3\!q\,\left\{{
G^2q_z^4\over(Gq_z^2 + E)^2}(\vec q_\perp\dot\vec u)^2  +
K_2q_z^2(q_\perp^2\delta_{ij}-q_iq_j)u_iu_j+K_3(q_z^2u)^2\right\}.}
One can see that if $Gq_z^2>E$ this action reduces to the infinite
polymer
action of \refs{\SB,\KLN}.  In other words, at short distance scales, even
finite polymers appear infinite.  The crossover to finite length
effects will occur when $Gq_z^2 = E$.  This corresponds to
a length scale
$\xi^L_z(\ell)= \sqrt{G/E}$.
At distances longer than this, the finite length effects
should become important.  At long distances though, finite polymers
appear only as points, and so we expect the system to
behave as a short molecule liquid crystal.

We would now like to see how the
the breakdown of conventional elasticity theory changes this length
scale.
The crossover to conventional elasticity still occurs at
the momentum at which
$E=Gq_z^2$.  This is modified from the harmonic result just derived
since $E$ becomes momentum
dependent itself.  {\sl A priori}, one might fear that $G$ acquires a
non-trivial momentum dependence as well;  however,
this does not happen, as the following argument shows:
The free energy \efreefull\ is invariant
under $\omega\rightarrow
\omega + h$ and $\vec u\rightarrow \vec u - {1\over d-1} h\vec x_\perp$.
Since the effective theory must have the same symmetry,
any graph that renormalizes {\sl any} quadratic
term in the free energy
$F$ involving $\omega$ must be absorbable into the renormalization
of the coefficient of $(\div\vec u + \omega)^2$
{\sl i.e.}, into the renormalization
of $E$. In other words, there are no graphs ``left over'' to renormalize $G$
which thus has only
trivial renormalizations.  Therefore we are able to find the
fluctuation corrected crossover length $\xi_\perp^L$.  Using
$E=E_0(q_\perp\xi_\perp^N)^{\eta_\perp/\zeta}$,
and converting $q_z$ to
$q_\perp$ by  $q_\perp\xi_\perp^N=(q_z\xi_z^N)^\zeta$, we have
\eqn\ethecross{\xi_\perp^L(\ell) = \xi_\perp^N\left[{G\over E_0(\xi_z^N
)^2}\right]^{\zeta\over 2-\eta_\perp} =
\xi_\perp^N\left(\ell\over\ell_0\right)^\gamma =
 \xi_\perp^N\left[{\ell a^2\over
2P^3}\right]^\gamma}
where $\gamma=0.67\pm 0.01$ and $\ell_0$ is given by \theellone .
The final equality holds only for steric interactions, although the
$\ell$ dependence is universal.

Replacing $q_\perp$ with $(\xi^L_\perp)^{-1}$ in the expressions \eanom\
for $K_{2,3}(\vec q)$ yields the length-dependent $K_{2,3}$
\Flen\ with $\phi_{2,3} = \eta_{2,3}/(1+\eta_2+\eta_3)$.
$K_1(\ell)$ can be obtained from
$K_1(\ell)={E\left(q_\perp=1/\xi^L_\perp, q_z=0\right) /(q^P_z(\ell))^2}$
\MEYER , with $q^P_z(\ell)$ obtained as described above.

To study the effect of chirality, we add the usual chiral term
\eqn\echiral{\eqalign{F_{\rm chiral} &= \int d^3\!x\,(c-c^*){\hat
n}\cdot\nabla\!\times\!
\hat n\cr &\approx \int d^3\!x\,\left\{(c-c^*)\left[1-(\dz\vec u)^2\right]
\curl\dz\vec u  - 2(c-c^*)\dz u_x\partial^2_z u_y\right\}\cr}}
where $c$ is proportional to
the concentration of a particular chiral species
and $c^*$ is the value of $c$ at which the handedness of the phase
changes from left to right.  In the second, approximate, equality, we
have done a few integrals by parts.

The pitch of the resulting phase is
gotten by balancing the twist term with the chiral term.  That is
$q^P=(c-c^*)/K_2$.  Hence we need the renormalized values of
$c$ and $K_2$.

We first note that the chiral term does not get renormalized
on length scales $<<P$, the pitch. On these length scales,
we are considering the renormalization group flow of
$(c-c^*)$ {\sl near} $(c-c^*)\approx 0$.
Fortunately,
the only graph that can renormalize the chiral term
is a tadpole which comes from the anharmonic term in
\echiral\ itself (the other terms in \efreefull\ have
chiral symmetry).
This is already proportional to $(c-c^*)$, so we
need not consider the correction to the propagator
coming from the quadratic term in \echiral\ to leading order in $c-c^*$.
Moreover,
this correction is simply proportional to
the real space mean squared director fluctuation
$\langle\,\dnb^2(\vec x)\,\rangle=\langle\,\vert\partial_z\vec u(\vec
x)\vert^2\,\rangle$
which is
finite in $d=3$. Thus $(c-c^*)$ acquires no divergent renormalizations.
This is in contrast to \PN , where the corrections were divergent
because, in membranes, $d=2$ and $\langle\,\dnb^2(\vec x)\,\rangle$
does diverge in that case, leading
to a non-trivial renormalization of $(c-c^*)$.

Finally, we calculate the pitch of polymer cholesterics,
given the chirality $(c-c^*)$.  In mean field theory
the pitch is found by balancing the twist term $K_2(\curl\dz\vec u)^2$
with the source term $(c-c^*)\curl\dz\vec u$.  The resulting
pitch is the inverse of
the momentum at which $\delta n(\vec q) = \bo{1}$, or, in other
words, $q_zu(\vec q)= \bo{1}$.  This gives the familiar mean field result
$q_\perp^P \sim (c-c^*)/K_2^0$ for the wavenumber $q_\perp^P
\identical {2 \pi / P}$ of the twisted structure, where P is its
pitch.
Since the source term
for twist is not renormalized and the effect of fluctuations can, by
construction, be absorbed into effective, momentum-dependent
Frank constants, we simply replace $K_2^0$ with $K_2(\vec q)$.  Now
balancing the twist and source in the effective theory, we find
\eqn\balance{
K_2(q_\perp)q_\perp^2[q_zu(\vec q)]^2 = (c-c^*)q_\perp [q_zu(\vec q)].}
Using
$K_2(q_\perp , q_z=0) =K^0_2 (q_\perp\xi^N_\perp)^{-\eta_2/\zeta} \times
\bo{1}$, $q_z u =\bo{1}$,
and $q_\perp^P \identical {2 \pi/ P}$
we find equation \pitpit\ of the introduction, with
$\nu ={\zeta\over \zeta-\eta_2}=
{2+\eta_2-\eta_3\over 2 -\eta_2-\eta_3}$,
where we have used the scaling
relation for $\zeta$.
Once the pitch is greater than
$\xi_\perp^L(\ell)$, it obeys the conventional, short molecule
cholesteric law $P\propto P_0$
because the elasticity is again conventional.
However, $P$ now depends on the polymer length
since $K_2$ does (as in \Flen). Using
\balance\ again to determine the pitch
yields the $\ell$ dependent pitch given in
equation \lpitch .

DNA provides an excellent experimental example of a polymer
cholesteric.
The DNA sample in
\PRP\ had an orientational
persistence length of $L_P=600\angstrom$ and mean polymer spacings
$a=35\angstrom$.  The typical length of their DNA molecules was
$4$ {\sl mm}!
Using our estimates for the
crossover lengths $\xi_\perp^N$ and $\xi_\perp^L$, we find that
$\xi_\perp^N\approx 1500\angstrom$ and $\xi_\perp^L\approx 4\times 10^4
\angstrom$, which leaves a respectable range of length scales ($1.4$
decades)
to observe the scaling behavior in \pitpit .
While the nonlinear effect is due to long molecules, the chirality of
the system could be changed by adding short molecules into the mix, if
they could be dissolved.  By cutting up the DNA with enzymes,
it should be possible
to move from the nonlinear to linear regime by shortening $\xi_\perp^L$,
and thereby observe the $\ell$ dependence \lpitch\ of the pitch.

The fact that DNA can have a rather long pitch of $\sim
10^4\angstrom$ \POD\
could be partially explained by our result.
Taking the above numbers, and a ``bare''
cholesteric pitch for DNA of $7000\angstrom$, equation \pitpit\
predicts that the actual pitch will be doubled from this bare value.

In some cases \ref\NPRIV{We thank David Nelson for calling this
fact to our attention.}\ DNA shows hexatic
order in the plane
orthogonal to $\hat n$. This ``N+6'' \ref\TONERii{This precise
type of orientational order was first proposed in J.~Toner, Phys.
Rev. A {\bf 27}, 1157 (1983).}\ structure leads to
an additional chiral coupling \KN , which
might tremendously increase the pitch.  This may explain why no twist
is observed in these systems \POD .
This is currently under investigation \KNT .

It is a pleasure to acknowledge stimulating discussions with T.~Lubensky,
D.~Nelson and R.~Podgornik.
RDK thanks IBM where some of this work was done; JT likewise
thanks the Isaac Newton Institute, Cambridge, UK.
RDK was supported in part by the National Science Foundation,
Grant
No.~PHY92--45317, and the Ambrose Monell Foundation.
\vfill\eject\immediate\closeout\ffile{\parindent40pt
\baselineskip.33truein\centerline{{\bf Figure Captions}}\nobreak\medskip
\escapechar=` \input figs.tmp\vfill\eject}
\footatend\vfill\supereject\immediate\closeout\rfile\writestoppt
\baselineskip=.33truein\centerline{{\bf References}}\bigskip{\frenchspacing%
\parindent=20pt\escapechar=` \input refs.tmp\vfill\eject}\nonfrenchspacing
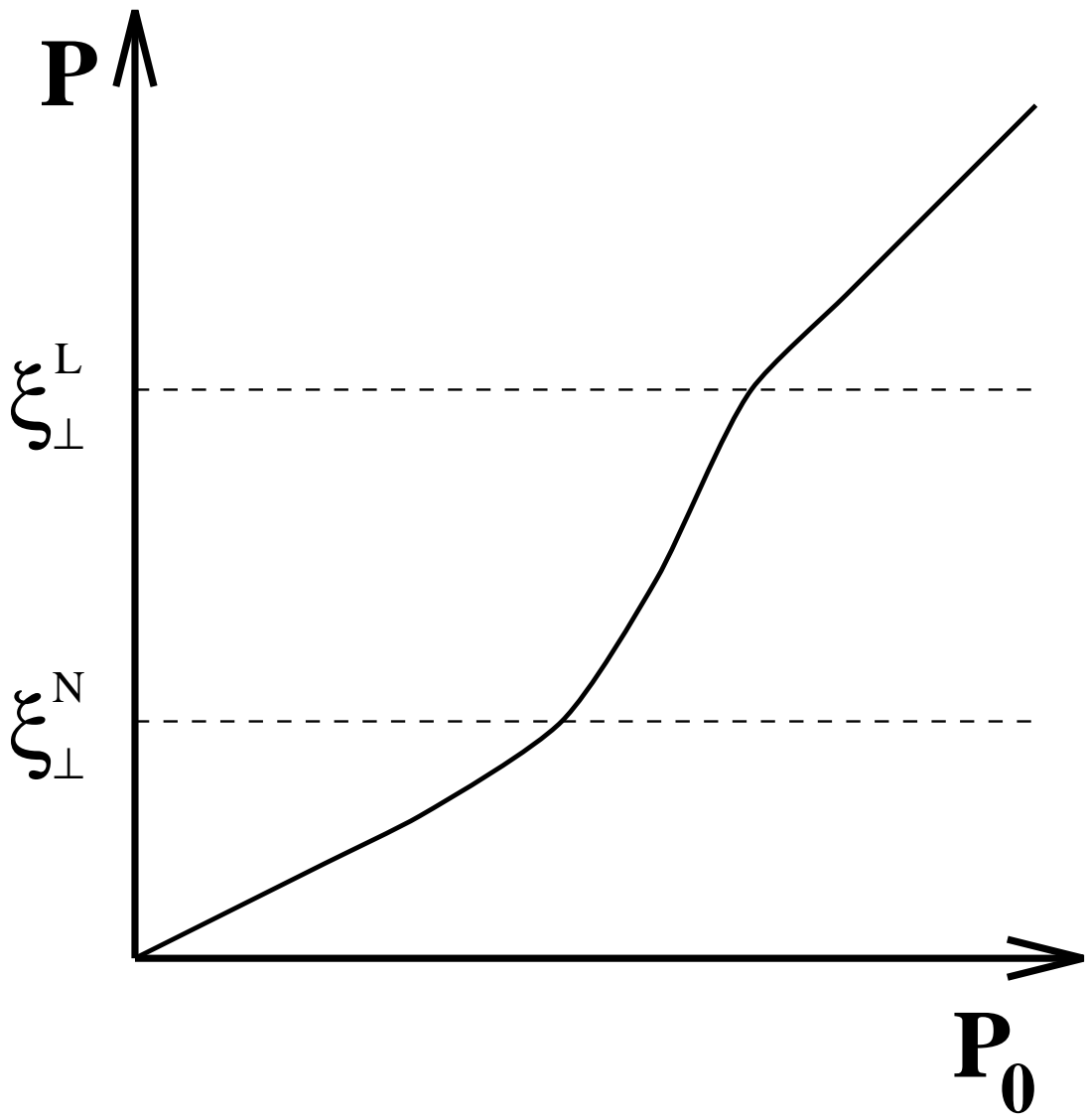
\bye